\documentclass[%
 aip,
 amsmath,amssymb,
 reprint,%
]{revtex4-1}

\usepackage{graphicx}
\usepackage{dcolumn}
\usepackage{bm}

\usepackage[utf8]{inputenc}
\usepackage[T1]{fontenc}
\usepackage{mathptmx}
\usepackage{etoolbox}
\usepackage[dvipsnames]{xcolor}
\usepackage{float}
\usepackage{color}

\usepackage[caption = false]{subfig}
\makeatletter
\def\@email#1#2{%
 \endgroup
 \patchcmd{\titleblock@produce}
  {\frontmatter@RRAPformat}
  {\frontmatter@RRAPformat{\produce@RRAP{*#1\href{mailto:#2}{#2}}}\frontmatter@RRAPformat}
  {}{}
}%
\makeatother
\begin{document}

\preprint{AIP/123-QED}
\title{The effect of nutation angle on the flow inside a precessing cylinder and its dynamo action}

\author{Vivaswat Kumar}
\altaffiliation{Author to whom correspondence should be addressed: v.kumar@hzdr.de}
\affiliation{Institute of Fluid Dynamics, Helmholtz-Zentrum Dresden-Rossendorf, Bautzner Landstr. 400, 01328 Dresden, Germany}

\author{Federico Pizzi}
\affiliation{Institute of Fluid Dynamics, Helmholtz-Zentrum Dresden-Rossendorf, Bautzner Landstr. 400, 01328 Dresden, Germany}
\affiliation{Department of Aerodynamics and Fluid Mechanics, Brandenburg University of Technology, Cottbus-Senftenberg, 03046 Cottbus, Germany}

\author{Andr\'e Giesecke}
\affiliation{Institute of Fluid Dynamics, Helmholtz-Zentrum Dresden-Rossendorf, Bautzner Landstr. 400, 01328 Dresden, Germany}

\author{J\'an \v{S}imkanin}
\affiliation{Institute of Geophysics of the Czech Academy of Sciences, Bo\v{c}n\'i II/1401, 141 31 Praha 4-Spo\v{r}ilov, Czech Republic}

\author{Thomas Gundrum}
\affiliation{Institute of Fluid Dynamics, Helmholtz-Zentrum Dresden-Rossendorf, Bautzner Landstr. 400, 01328 Dresden, Germany}

\author{Matthias Ratajczak}
\affiliation{Institute of Fluid Dynamics, Helmholtz-Zentrum Dresden-Rossendorf, Bautzner Landstr. 400, 01328 Dresden, Germany}

\author{Frank Stefani}
\affiliation{Institute of Fluid Dynamics, Helmholtz-Zentrum Dresden-Rossendorf, Bautzner Landstr. 400, 01328 Dresden, Germany}

\date{\today}
\begin{abstract}
The effect of the nutation angle on the flow inside a precessing cylinder is experimentally explored and compared with numerical simulations. 
The focus is laid on the typical breakdown of the directly forced $m=1$ Kelvin mode for increasing precession ratio (Poincaré number), and the accompanying transition between a laminar and turbulent flow.
Compared to the reference case with a 90$^{\circ}$ nutation angle, prograde rotation leads to an earlier breakdown, 
while in the retrograde case the forced mode continues to exist also
for higher Poincar\'e numbers. Depending largely on the  occurrence and intensity of an axisymmetric double-roll mode, a kinematic dynamo study reveals a sensitive dependency of the self-excitation condition on the nutation angle and the Poincaré number. Optimal dynamo conditions are found for 90$^{\circ}$ angle which, however, might shift to slightly retrograde precession for higher Reynolds numbers.
\end{abstract}

\maketitle



\section{Introduction}
Precession-induced fluid motion is essential in various phenomena and applications, including fuel payloads of rotating spacecrafts, atmospheric vortices like tornadoes and hurricanes, and the flow in the Earth's liquid outer core. The suggestion that precessional forcing can act as a potential power source for Earth's magnetic field \cite{bullard1949} had initiated a debate on the non-trivial issue of the energy budget required for the geodynamo. The early argument that a precessing laminar flow cannot convert enough energy to maintain Earth's magnetic field \cite{loper1975, rochester1975} changes completely for turbulent flows which can dissipate more energy, thereby making it possible to sustain the geomagnetic field \cite{malkus1968precession}. Meanwhile precession is also believed to be responsible for the dynamos of the ancient moon \cite{dwyer2011, lebars2011} and the asteroid Vesta \cite{fu2012}, and several numerical studies have evidenced that magnetic fields indeed can be generated via precession-driven flows \cite{tilgner2005precession, nore2011nonlinear, wu2013dynamo}. 

In the meantime, a variety of experiments focusing on precession-driven flows were conducted in numerous laboratories \cite{mcewan1970inertial, vanyo1971measurement, manasseh1992breakdown, kobine1995inertial, noir2001experimental, goto2007turbulence, mouhali2012evidence, herault2015subcritical}.
The closest to a hydromagnetic dynamo was that of Gans\cite{gans1971hydromagnetic} who performed a precession-driven liquid metal experiment in 1971, and achieved a magnetic field amplification by a factor of 3. This experiment motivated the development of a large-scale precession dynamo experiment, which is presently under construction at Helmholtz-Zentrum Dresden-Rossendorf (HZDR) within the framework of the DRESDYN project \cite{stefani2019dresdyn}. 
After the successes of the pioneering liquid sodium experiments
in Riga \cite{gailitis2018self}, Karlsruhe\cite{muller2006experiments}, and Cadarache \cite{monchaux2009karman}, the DRESDYN experiment aims at achieving truly homogeneous dynamo action without the use of any propellers, pumps, guiding tubes or magnetic materials.
Prior studies had indeed demonstrated that precession can act as a efficient mechanism for driving an intense flow in a homogeneous fluid \cite{leorat2006large}.

The DRESDYN precession experiment consists of a cylinder with a radius of R = 1 m and a height of H = 2 m. The cylinder rotates around its symmetry axis at the cylinder frequency $f_c \leq$ 10 Hz and precesses around another axis at the precession frequency $f_p \leq$ 1 Hz. In this experiment, liquid sodium will be used as a working fluid to accomplish dynamo action \cite{stefani2019dresdyn, giesecke2019kinematic}. To understand the dynamics of the flow for the large-scale experiment, a 1:6 down-scaled water test experiment with the same aspect ratio and rotation rates has been in operation at HZDR for many years \cite{herault2015subcritical, herault2019, giesecke2018nonlinear}.

Precession produces complex three-dimensional flow structures as a consequence of the interaction between inertial modes, boundary layers, and the directly driven base flow \cite{giesecke2019kinematic, herault2015subcritical, pizzi2021prograde}. In cylindrical geometry, the inertial modes interact to form global modes known as Kelvin modes \cite{kelvin1880vibrations}. Each mode has its own eigen-frequency, which is determined by the radial ($n$), axial ($k$), and azimuthal ($m$) wave numbers. The Reynolds number $Re$, the precession ratio or Poincare number ($Po=\Omega_p/\Omega_c$), the geometric aspect ratio ($\Gamma =$ H/R), and the nutation angle $ \pm \alpha$ (angle between the axis of precession and the axis of rotation) determine the flow inside the precessing cylinder. Here, $+\alpha$ represents prograde precession and $-\alpha$ represents retrograde precession (prograde/retrograde precession occurs when the projection of the turntable rotation on the cylinder rotation is positive/negative). 

The present study examines the influence of different nutation angles on the dominant flow modes inside the precessing cylinder for these rotation configurations. For this purpose we conduct water experiments and compare their results with numerical simulations\cite{pizzi2021prograde}. Focusing on the inertial modes with the largest energy fractions we use direct measurements of the axial velocity with ultrasound Doppler velocimetry (UDV) and simulation data from a three dimensional numerical model. Finally, we utilize the obtained flow fields in a kinematic dynamo model, in order to identify the most promising parameter range for dynamo action in the upcoming DRESDYN experiment. In this kinematic dynamo model the flow is prescribed using the full information obtained from numerical simulations of the hydrodynamic problem, whereas at this stage any feedback of the field on the flow via the Lorentz force is still ignored.

The paper is structured as follows: In section 2, we briefly describe the setup and the employed measurement procedure for the down-scaled water experiment. The theory and numerics is described in section 3, including  both the hydrodynamic and the kinematic dynamo problem. Our findings from hydrodynamic 
experiments and simulations are described and compared in section 4.  The results on dynamo action for different configurations are then discussed in section 5. In the final section, the results are summarized and some prospects for future work and the large-scale liquid sodium experiment are discussed.

\section{Down-scaled water precession experiment}
\subsection{Experiment setup}
Figure 1(a) gives a schematic view of the 1:6 down-scaled water precession experiment. The experiment comprises a water-filled acrylic cylindrical vessel with radius $R = 163$\, mm and height $H = 326$\, mm. The vessel is connected with an asynchronous 3 kW motor via a transmission chain, which allows to adjust the rotation rate of the cylinder. The rotational frequency of the cylinder $f_c$ can reach a maximum of 10 Hz. The cylinder's end caps are joined axially by eight rods to keep their alignment in parallel, as shown in Fig. 1(a). This entire system is mounted on a turntable powered by a second 2.2 kW asynchronous motor, which can rotate up to a frequency $f_p$ of 1 Hz. Both rotation rates, i.e. $f_c$ and $f_p$, are continuously measured by two tachometers and recorded by a data acquisition system.
The nutation angle $\alpha$ formed by the cylinder rotation axis and the turntable rotation axis can be varied between 60$^{\circ}$ and 90$^{\circ}$. The present study conducts experiments for the three different nutation angles of $\alpha$ = 60$^{\circ}$, 75$^{\circ}$ and 90$^{\circ}$. 
\begin{figure}[h!]
\centering
\includegraphics[width=0.35\textwidth]{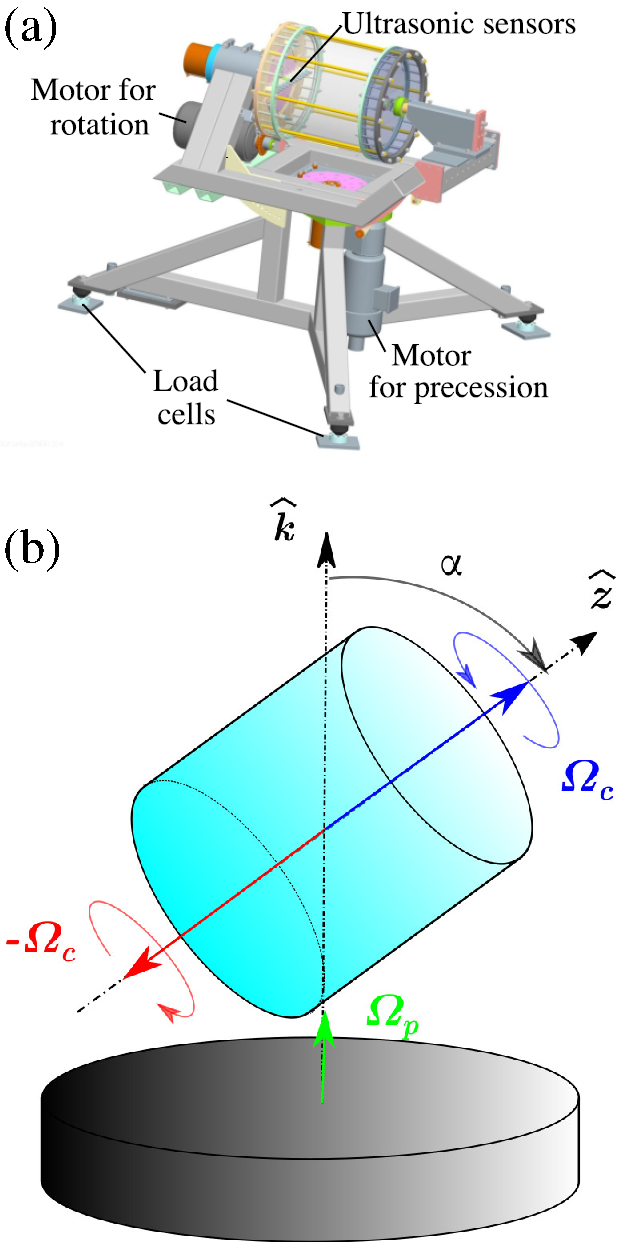}
\caption{(a) Schematic setup of the 1:6 down-scaled water precession experiment; 
(b) Sketch of the precessing cylinder with the container $\Omega_{c}=2 \pi f_c$ and the precession angular velocity $\Omega_{p}=2 \pi f_p$ . Here $\alpha$ is the nutation angle.
\cite{stefani2012dresdyn, pizzi2021prograde}.
}
\label{fig:cylinder_sketch}
\end{figure}

\subsection{Measurement procedure}
To determine the flow field inside the cylinder, ultrasonic transducers (TR0408SS, Signal Processing SA, Lausanne) are placed at one end cap of the cylinder (Fig. 1(a)). These transducers are connected with an ultrasound Doppler velocimeter (UDV) (Dop 3010, Signal Processing SA, Lausanne), which records the velocity profiles with a temporal resolution of about 10 Hz. Each transducer emits an ultrasonic pulse and receives echoes reflected from particles in the path of the ultrasonic beam on a regular basis. The velocimeter then infers the axial flow velocity in front of the sensor from the Doppler shift of the recorded echoes.
The tracer particles used in the water inside the cylindrical vessel consist of a mixture of Griltex 2A P1 particles with sizes of 50 $\mu$m (60 \% by weight) and 80 $\mu$m (40 \% by weight). These tracer particles have a density of 1.02 g/cm$^3$. Before commencing the experimental run, the water is vigorously mixed at a high rotation rate to ensure that the tracer particles are distributed evenly throughout the cylindrical vessel. The vessel is then set into rotation at $f_c$ until the fluid co-rotates with the vessel, which is indicated by a vanishing velocity on the UDV channel. We then set the turntable in motion at $f_p$ and wait until a statistically steady state is attained. The velocity profile is then recorded for approximately 52 rotations of the cylinder. Finally, we increase the precession rate to the next value of $f_p$ and wait until the fluid motion at the increased precession rate has reached a steady state.

Initially, the measurement started at a value of $Po = 0.035$ (i.e. $f_p = 0.002$\, Hz), which was then gradually increased in steps up to $Po = 0.198$ (i.e. $f_p = 0.011$\,Hz). The flow measurements were conducted with the ultrasonic transducer at radius $r = 150$\,mm and at a constant rotation rate of the cylinder $f_c = 0.058$\, Hz, corresponding to $Re \approx 10^4$. These experiments were carried out with two different rotation directions, i.e. prograde and retrograde precession. During the experiments, the room temperature was kept as constant as possible using air-conditioning, thereby eliminating the temperature-dependent viscosity as a possible influence on the flow.

\section{Theory and numerics}\label{sec:2}
\subsection{Hydrodynamics}

We consider an incompressible fluid of kinematic viscosity $\nu$ enclosed in a cylinder of radius $R$  and height $H$. The container rotates and precesses with angular velocities $\boldsymbol{\Omega_{c}}$ and $\boldsymbol{\Omega_{p}}$ ($-\boldsymbol{\Omega_{p}}$) for prograde (retrograde) motion, with $\alpha$ denoting the nutation angle, as illustrated in Fig. \ref{fig:cylinder_sketch}(b). Another option is considering $\alpha$ to run between $0^{\circ}$ and $180^{\circ}$, however we use here the range between $0^{\circ}$ and $90^{\circ}$, and distinguish between pro- and retrograde precession.\\
The fluid motion inside the precessing cylinder is governed by the Navier-Stokes equation 
\begin{eqnarray}\label{eq:dimensionless_ns}
\frac{\partial \boldsymbol{u}}{\partial t} + \boldsymbol{u} \cdot \boldsymbol{\nabla u}   = - \boldsymbol{\nabla}P + \frac{1}{Re} \boldsymbol{\nabla}^{2}\boldsymbol{u} - 2 \boldsymbol{\Omega} \times \boldsymbol{u} + \frac{d\boldsymbol{\Omega}}{dt} \times \boldsymbol{r} \:, 
\end{eqnarray}
together with the incompressibility condition $\boldsymbol{\nabla} \cdot \boldsymbol{u} = 0$. Here $\boldsymbol{u}$ is the velocity flow field, $\boldsymbol{\Omega}=\boldsymbol{\Omega_{c}}+\boldsymbol{\Omega_{p}}$ is the total rotation vector and $\boldsymbol{r}$ is the position vector with respect to the center of the cylinder. $P$ is the reduced pressure which comprises the hydrostatic pressure and other gradient terms, e.g. the centrifugal force, that do not change the dynamical behavior of the flow. The last two terms on the right-hand side are  the Coriolis and the Poincar\'{e} forces, respectively. The cylindrical coordinates are the axial ($z$), radial ($r$) and azimuthal ($\varphi$) ones, respectively. Equation~(\ref{eq:dimensionless_ns}) is complemented by no-slip boundary conditions at the walls. More specifically the axial $u_{z}$ and azimuthal $u_{\varphi}$ velocities vanish on the sidewall while on the endwalls the radial $u_{r}$ and azimuthal velocities are zero.\\
In order to non-dimensionalize the Navier-Stokes equation we use the radius $R$ as length scale and $\left| \Omega_{c}+\Omega_{p}\cos \alpha \right|^{-1}$ as time scale. The latter choice relies on using the projection of total angular velocity on the cylinder axis, i.e $(\boldsymbol{\Omega_{c}}+\boldsymbol{\Omega_{p}}) \cdot \boldsymbol{\widehat{z}}$.\\
The key dimensionless parameters governing precession-driven flows, the Reynolds number $Re$, the Poincar\'{e} number $Po$ and the aspect ratio of the container $\Gamma$, are defined as
\begin{eqnarray}
Re = \frac{R^{2}\left| \Omega_{c}+\Omega_{p}\cos \alpha\right|}{\nu}, \quad Po=\frac{\Omega_{p}}{\Omega_{c}}, \quad \Gamma=\frac{\textrm{H}}{\textrm{R}} \: .
\end{eqnarray}\label{groups_dimensionless}

\subsection{Magnetohydrodynamics}
The governing equation for the spatio-temporal evolution of the magnetic field $\boldsymbol{B}$ is the (dimensionless) induction equation:

\begin{eqnarray}\label{eq:induction}
\frac{\partial \boldsymbol{B}}{\partial t} = \boldsymbol{\nabla} \times \left( \langle \boldsymbol{u} \rangle \times \boldsymbol{B} - \frac{\nabla \times \boldsymbol{B}}{Rm} \right) \: .
\end{eqnarray}

For the time-averaged velocity field
$\langle \boldsymbol{u} \rangle$ considered
constant, the solution of the linear evolution equation has the form $\boldsymbol{B}=\boldsymbol{B}_{0} \exp \left( \sigma \: t\right)$ with $\sigma$ representing the eigenvalue. In Eq. (3) $Rm$ is the magnetic Reynolds number, defined as:
\begin{equation}
Rm=\frac{R^{2}\left| \Omega_{c}+\Omega_{p}\cos \alpha\right|}{\eta},
\end{equation}
where $\eta$ is the magnetic diffusivity of the liquid metal.


\begin{figure*}[!]
\centering
\includegraphics[width=0.9\textwidth]{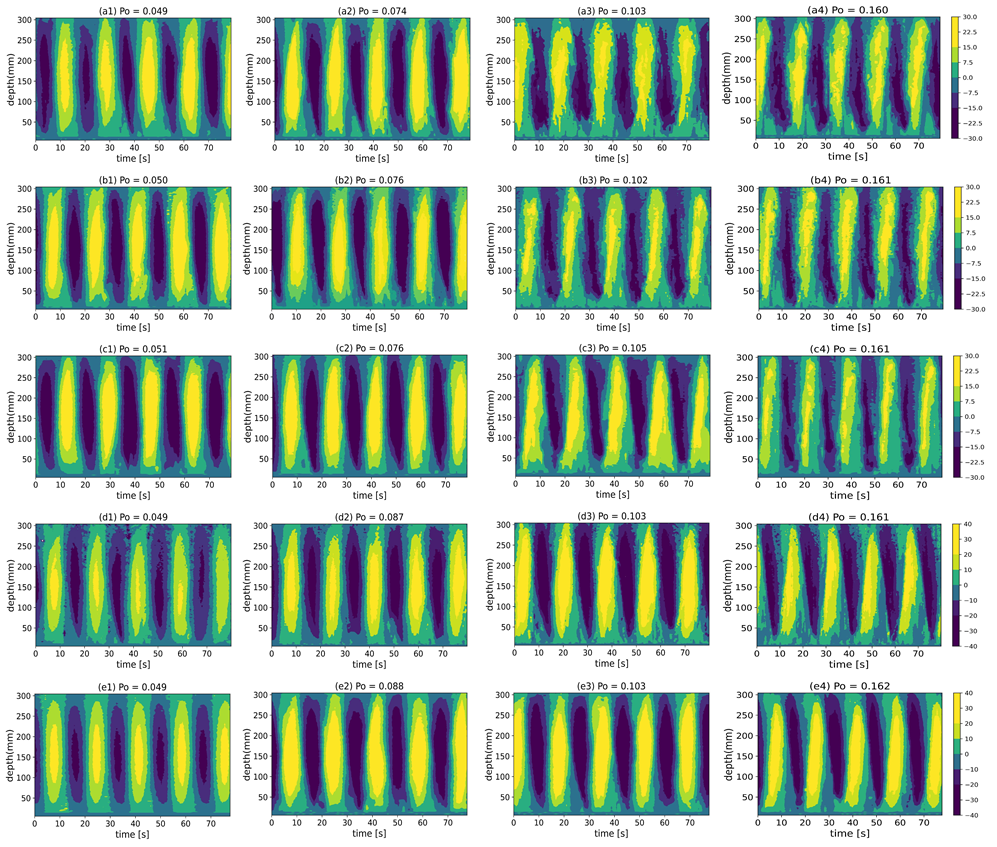}
\caption{From left to right: velocity contours over time and depth for increasing Po (appr. between 0.049 and 0.16). First row (a1-a4) for $\alpha= 60^{\circ}$ (prograde); Second row (b1-b4) for $\alpha= 75^{\circ}$ (prograde); Third row (c1-c4) for $\alpha= 90^{\circ}$; Fourth row (d1-d4) for $\alpha= 75^{\circ}$ (retrograde); Fifth row (e1-e4) for $\alpha= 60^{\circ}$ (retrograde) at $Re \approx 10^4$.}\label{fig:foobar}
\end{figure*}

\subsection{Numerical methods}
Precession-driven flows in cylinders can be simulated in one of the following two frames of reference: ($i$) the mantle frame (attached to the cylinder wall) or ($ii$) the turntable frame in which the cylinder walls rotate at $\Omega_{c}$ and the total vector $\Omega$ is fixed. We select the former for which $\partial \Omega/\partial t = 0$ so that the Poincare force disappears and both the rotation vector $\Omega_{c}$ as well the precession vector $\Omega_{p}$ become stationary. While for the hydrodynamic problem we use a spectral element-Fourier code \cite{semtex} (meshing the domain 300 quadrilateral elements to mesh the meridional half plane and 128 Fourier modes in azimuthal direction) with no slip boundary condition, we solve the induction equation through a finite volume scheme with constraint transport in order to ensure $\nabla \cdot \boldsymbol{B}=0$. For the dynamo simulations we apply pseudo-vacuum boundary condition for the magnetic field (only tangential components vanish at the wall).\\
The simulation protocol is the following: we start at $t=0$ with a pure solid body rotation motion i.e $\boldsymbol{u} = (\Omega_{c} \: r) \widehat{\boldsymbol{\varphi}}$ imposing a certain forcing. Once the statistically steady regime is achieved by the hydrodynamic flow field we average in time and we put this flow structure in the induction equation Eq.~(\ref{eq:induction}). The kinematic dynamo simulations are run till a diffusion time $t_{b} \geq Rm$.\\
The phase space investigated for the kinematic dynamo problem ranges between $Re \in [1000,10000] $  and the Poincar\'{e} number in the range $\pm[0.010 , 0.20]$. The nutation angle is in between $\alpha \in [60^{\circ},  75^{\circ}]$, both for prograde and retrograde precession. The aspect ratio will be fixed at $\Gamma = 2$, quite close to the resonance point $\Gamma = 1.989$ of the first inertial mode. The collection of all simulations in the parameter space $(Re, Po)$ will be later shown in Figs. \ref{fig:phase_diagrams} and \ref{fig:regime_re}.

\section{Results}
\subsection{Typical flow patterns}
In this subsection, we present the results of the precession water experiment. The contour plots of the flow structures obtained during the experiment are shown in Fig. \ref{fig:foobar}. These plots illustrate the evolution of the axial velocity profile $u_z$ over time $t$ and depth $z$ (the depth indicates the distance along the transducer's axis from the transducer). The first to fifth rows show the results for $\alpha$ = 60$^{\circ}$, 75$^{\circ}$ and 90$^{\circ}$ respectively, for both prograde and retrograde precession. The dominating oscillatory pattern of the velocity profile, defined by the rotational frequency $\Omega_c$ of the cylinder, represents the standing inertial mode with $(m,k)=(1,1)$ as recorded by the rotating UDV sensor mounted on the vessel wall.

As we increase  $f_p$, the response of the fluid begins to change. At lower values of 
$Po$ 
(see Fig. \ref{fig:foobar}, first column), the flow pattern exhibits a stable flow structure and is almost vertical in both cases (prograde and retrograde). However, as $Po$ exceeds a certain higher value for $\alpha$ = 60$^{\circ}$ (prograde), $\alpha$ = 75$^{\circ}$ (prograde) and 90$^{\circ}$, as shown in the third column of Fig. \ref{fig:foobar} (a3, b3 and c3), the flow pattern changes and exhibits a significant tilt with respect to the vertical axis. By contrast, in the retrograde case (see Fig. \ref{fig:foobar} (d3) and (e3)), it remains almost unchanged up to this $Po$ value, and the occurrence of the tilt is shifted to a higher value of $Po$. The tilt indicates a flow state transition at a critical value of the precession ratio ($Po^c$), implying the presence of other inertial modes \cite{giesecke2019kinematic}. In addition, the transition that occurred at the critical $Po$ has a considerable effect on the amplitudes of the inertial modes, as we will demonstrate in the following.

\subsection{Quantitative results and comparison with numerics}

In a more quantitative analysis, the mode amplitudes are calculated by decomposing the measured axial velocity field in the axial and azimuthal directions (for a comprehensive overview of the calculation of the amplitudes of inertial modes, we refer the reader to a publication \cite{giesecke2019kinematic}). While, in principle, various modes can be observed inside the precessing cylinder, in our analysis we examined only those modes that have substantial amplitudes and are most relevant for dynamo action\cite{giesecke2018nonlinear}, i.e., $(m, k) = (1, 1)$ and 
$(m, k) = (0, 2)$.

\begin{figure}[!]
\centering
\includegraphics[width=0.45\textwidth]{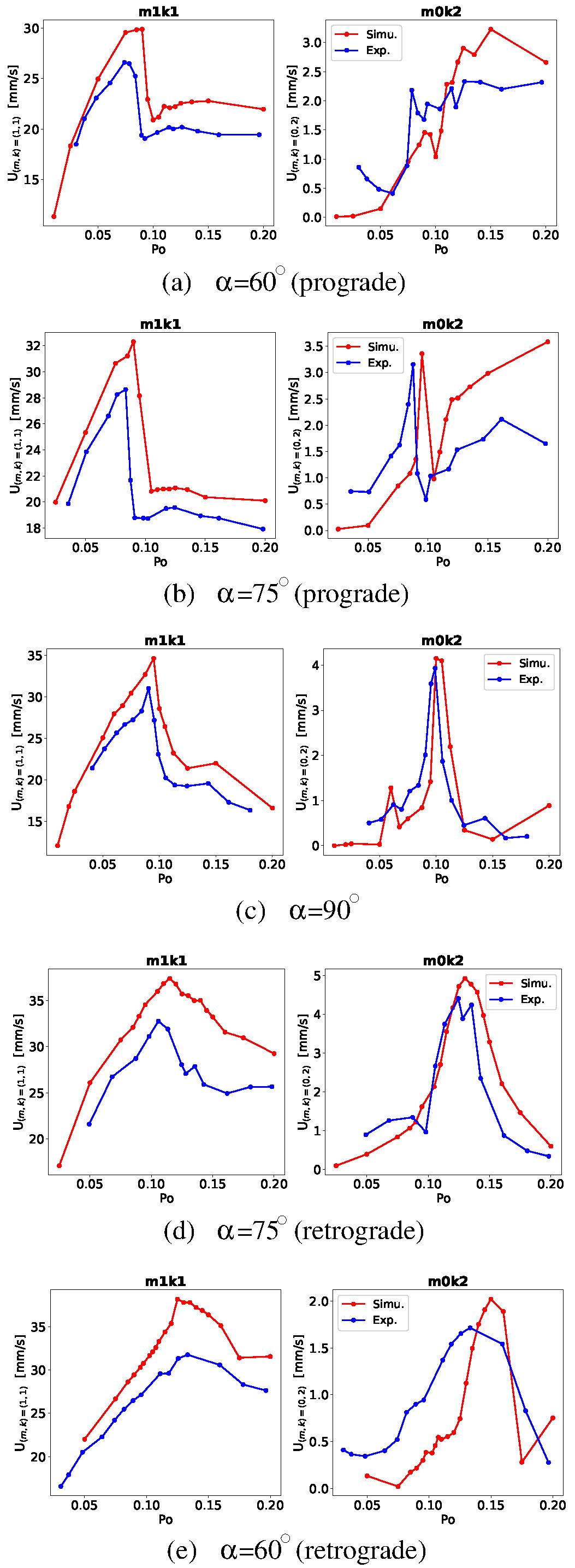} 
\caption{Comparison of amplitudes of the directly forced mode $(m, k) = (1, 1)$ and the axisymmetric mode $(m, k) = (0, 2)$ between numerically calculated (red) and experimentally measured flow (blue) for a nutation angle of 60$^{\circ}$, 75$^{\circ}$ and 90$^{\circ}$.\label{fig:compare}}
\label{}
\end{figure}

Figure \ref{fig:compare} shows the amplitudes of the prominent modes versus the precession ratio for both cases (prograde and retrograde) at 60$^{\circ}$, 75$^{\circ}$, and 90$^{\circ}$, respectively. In order to compare with the experimental data, the simulation results at $Re = 6500$ were linearly extrapolated to the experimental value $Re = 10^4$. For prograde 60$^{\circ}$, 75$^{\circ}$ and 90$^{\circ}$ (see Fig. \ref{fig:compare}(a), \ref{fig:compare}(b) and \ref{fig:compare}(c)), we observe that the directly forced mode $(m, k) = (1, 1)$ rises up to $Po \approx 0.08$, beyond which there is an abrupt transition of the flow state due to the breakdown of the $(m, k) = (1, 1)$ mode. Simultaneously, an axially symmetric mode $(m, k) = (0, 2)$ appears in a narrow range of $Po$, which corresponds to the double roll structure that was previously shown to be most relevant for dynamo action \cite{giesecke2019kinematic}. Remarkably, the nutation angle influences the critical $Po$, such that as the angle increases (60$^{\circ}$, 75$^{\circ}$ and 90$^{\circ}$), so does the critical $Po$ (0.083, 0.087 and 0.01). In other words, for smaller nutation angles the transition sets in earlier.

In contrast, the data for the retrograde $60^{\circ}$ and $75^{\circ}$ cases show no clear breakdown of the directly forced mode $(m, k) = (1, 1)$ which has a gradual decrease in amplitude. At the same time, we observe a smoother increase of the axially symmetric mode $(m, k) = (0, 2)$ within the considered range of $Po$, as shown in Figs. \ref{fig:compare}(d) and \ref{fig:compare}(e). In comparison to all other cases, $\alpha = 75^{\circ}$ (retrograde) acquires the largest amplitude of the $(m, k) = (0, 2)$ mode, and $\alpha = 60^{\circ}$ (retrograde) has the smallest. As compared to the prograde case the critical Po values are shifted to larger values for retrograde case. The slight offsets of the experimental data with respect to the numerical data along the x-axis is probably due to the difference in Re: as Re increases, its critical $Po$ decreases slightly\cite{pizzi2021prograde}. In general, however, the experimental values at  $\alpha = 60^{\circ}, 75^{\circ}$ for two different configurations (prograde and retrograde) and at $90^{\circ}$ are in good agreement with the results of the numerical simulations. 



\section{Dynamo results}\label{sec:main}
In this section we present the results of the kinematic dynamo code applied to the flow fields as obtained in the previous section. The analysis will focus on two main points: $(i)$ the influence of the nutation angle $\alpha$ on the ability to drive dynamo action; $(ii)$ the impact of the Reynolds number for a fixed angle $\alpha=90^{\circ}$.

\begin{figure}[h!]
\centering
\includegraphics[width=0.45\textwidth]{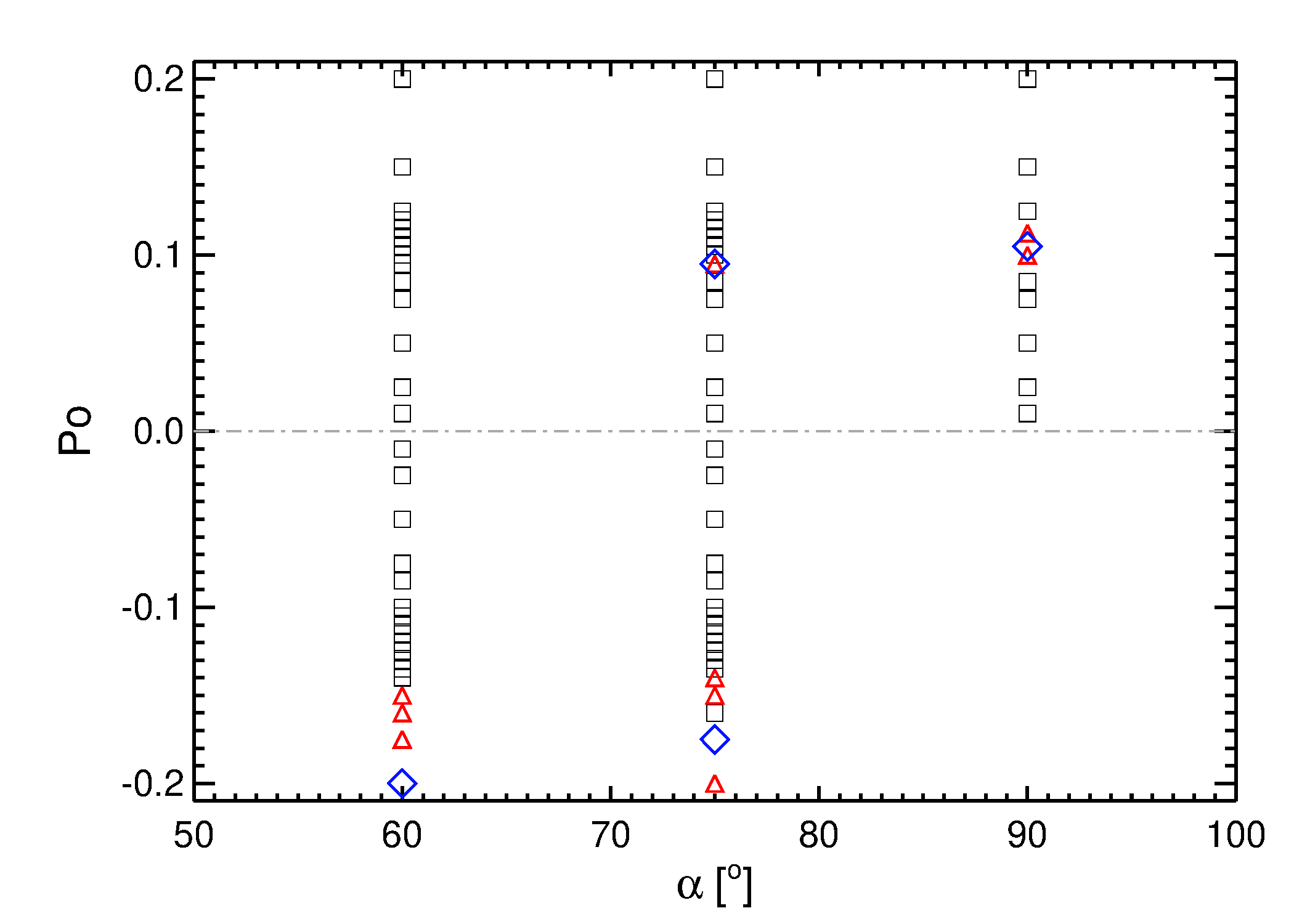}
\caption{Regime diagram of the kinematic dynamo simulations in the $(Po,\alpha)$ parameter space with fixed $Re=6500$. Prograde (retrograde) cases correspond to the positive (negative) region $Po>0$ ($Po<0$). Red symbols indicate dynamo action; black symbols show missing dynamo action and the blue diamond show the peak of strongest dynamo for each angle.\label{fig:phase_diagrams}}
\end{figure}

\begin{figure}[!]
\centering
\includegraphics[width=0.45\textwidth]{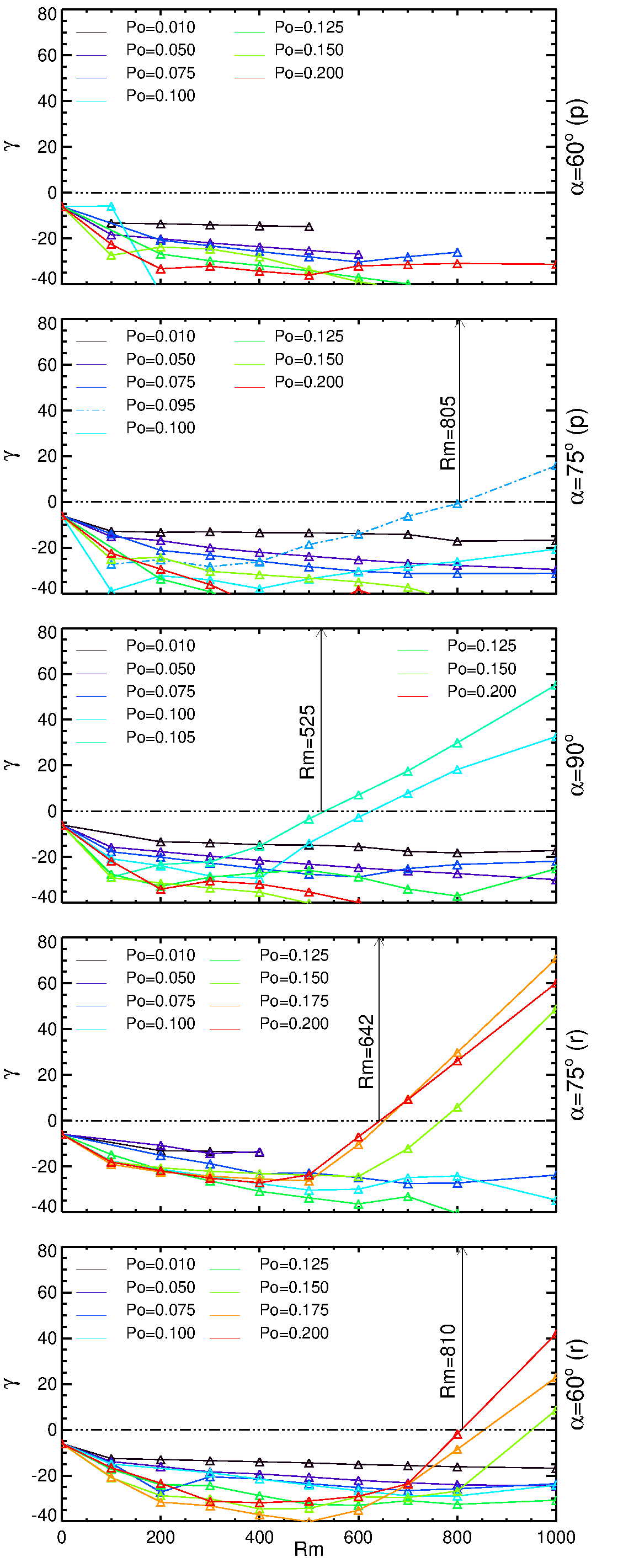}
\caption{Growth rate $\gamma$ of the magnetic energy as a function of the magnetic Reynolds number for five nutation angles. Various curves represent the different precession ratios and the arrows mark the dynamo onset at the critical $Rm$. \label{fig:growth_rate}}
\end{figure}

\subsection{The role of the nutation angle}\label{subsec:alpha_role}
\begin{figure*}[!]
\centering
\includegraphics[width=0.9\textwidth]{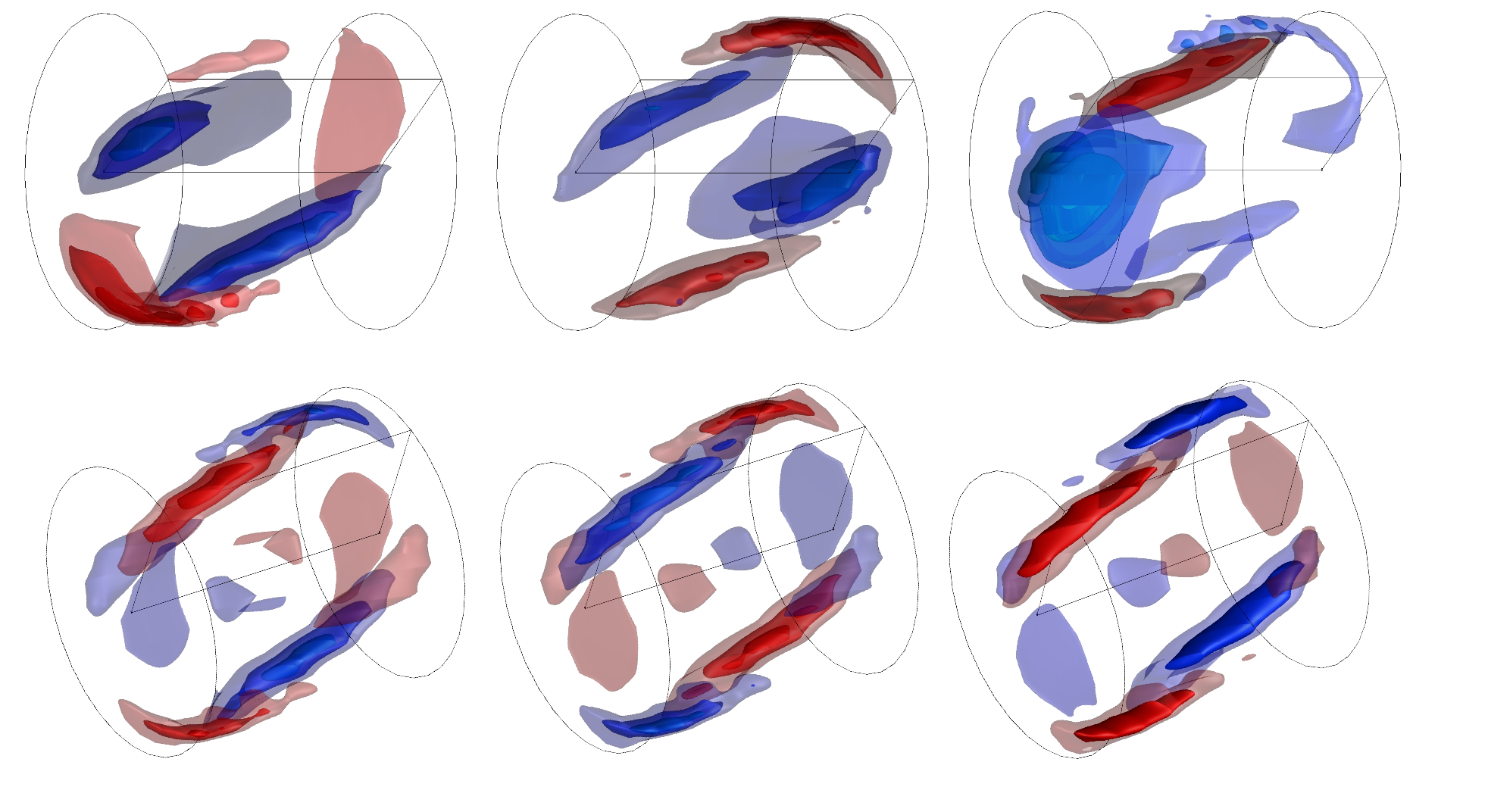}
\caption{Three snapshots of the azimuthal magnetic field $B_{\varphi}$ at $Rm=700$: top row $\alpha=90^{\circ}$ and $Po=0.105$ ; bottom row : $\alpha=75^{\circ}$ retrograde and $Po=0.175$. Blue colour denotes negative values and red colour positive with the levels of translucency denoting $30\%$, $50\%$, $70\%$ of the field. \label{fig:snapshots}}
\end{figure*}

\begin{figure}[h!]
\centering
\includegraphics[width=0.45\textwidth]{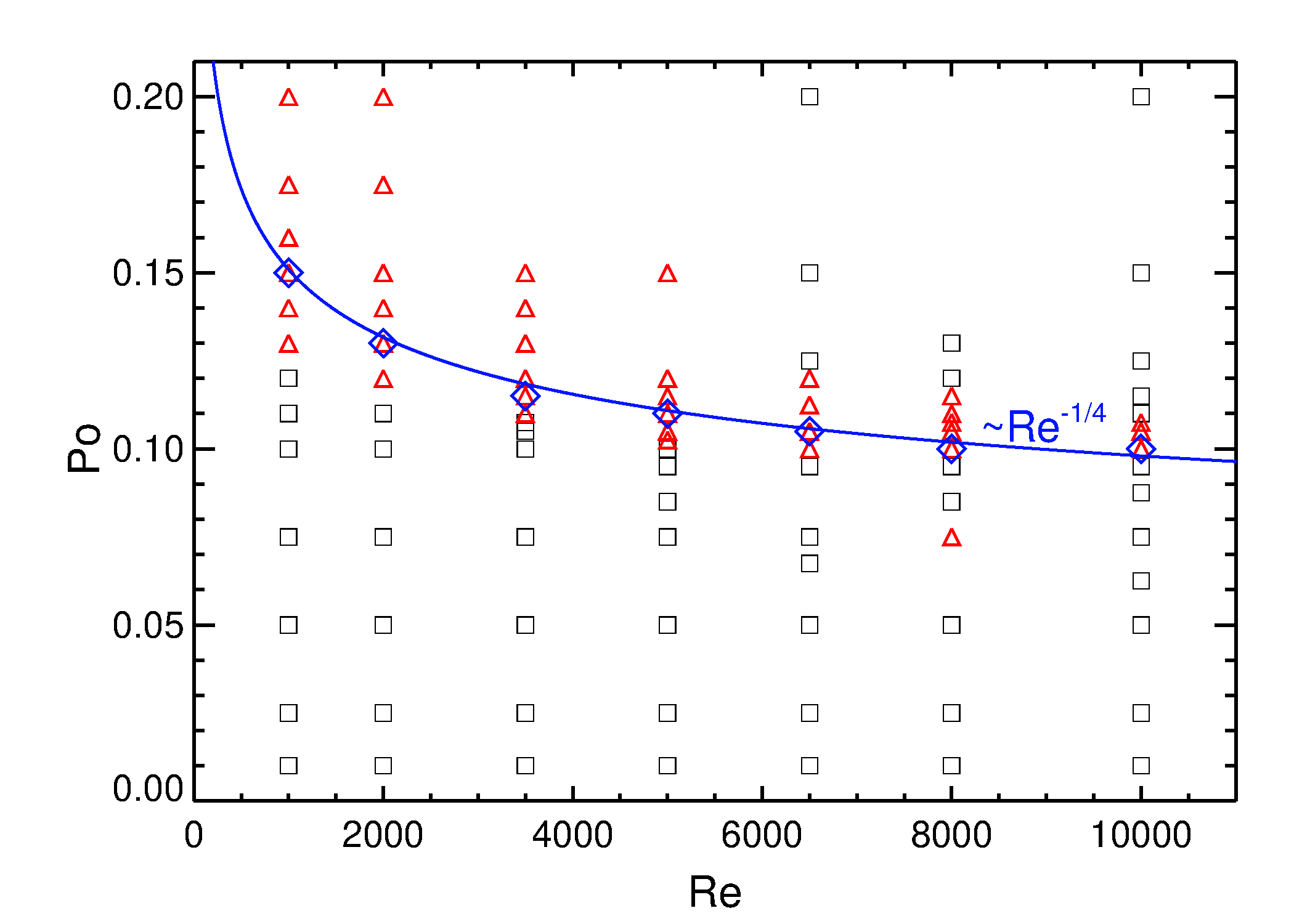} 
\caption{$(Po, Re)$ parameter space with fixed $\alpha=90^{\circ}$. Black symbols represent no dynamo effect and red symbol dynamo action found in the range of $0 < Rm \leq 10^{3}$. Blue diamonds highlight the best dynamo action for each $Re$ and the corresponding blue line is the scaling law $Po^{(c)} \sim Re^{-1/4}$. \label{fig:regime_re}}
\end{figure}

Figure~\ref{fig:phase_diagrams} shows the regime diagram in the $(\alpha, Po)$ space at fixed $Re=6500$. We find dynamo action (red symbols) for all the angles except for $\alpha=60^{\circ}$ prograde. The range of the precession ratio where dynamo action occurs changes with the nutation angle: for prograde cases dynamos occur at $Po \approx 0.1$ while for retrograde they appear at $Po > 0.15$ with a more extended range. For each angle, the blue diamonds indicate the dynamo with the largest growth rate.
\\
We plot the growth rate $\gamma= 2\: \Re \left( \sigma \right)$ of the magnetic field  in Fig.~\ref{fig:growth_rate}. As already highlighted,  the $\alpha=60^{\circ}$ prograde shows no positive growth rate even at the largest magnetic Reynolds number considered here. The lowest critical magnetic Reynolds number occurs for $\alpha=90^{\circ}$ which, therefore, turns out to be the most promising case for the later dynamo experiment.\\
Also the magnetic field structure, in this case the azimuthal component $B_{\varphi}$, depends on $\alpha$ (Fig.~\ref{fig:snapshots}). The three snapshots are taken between $t=300$ and and $t=380$. Both cases present contours elongated along the axis and the final field shows a change in sign during the evolution. This feature could be either a rotation or reversal in the sense of `active longitudes' that can be found in some astrophysical objects.

\subsection{The role of Reynolds number}
\begin{figure}[h!]
\includegraphics[width=0.45\textwidth]{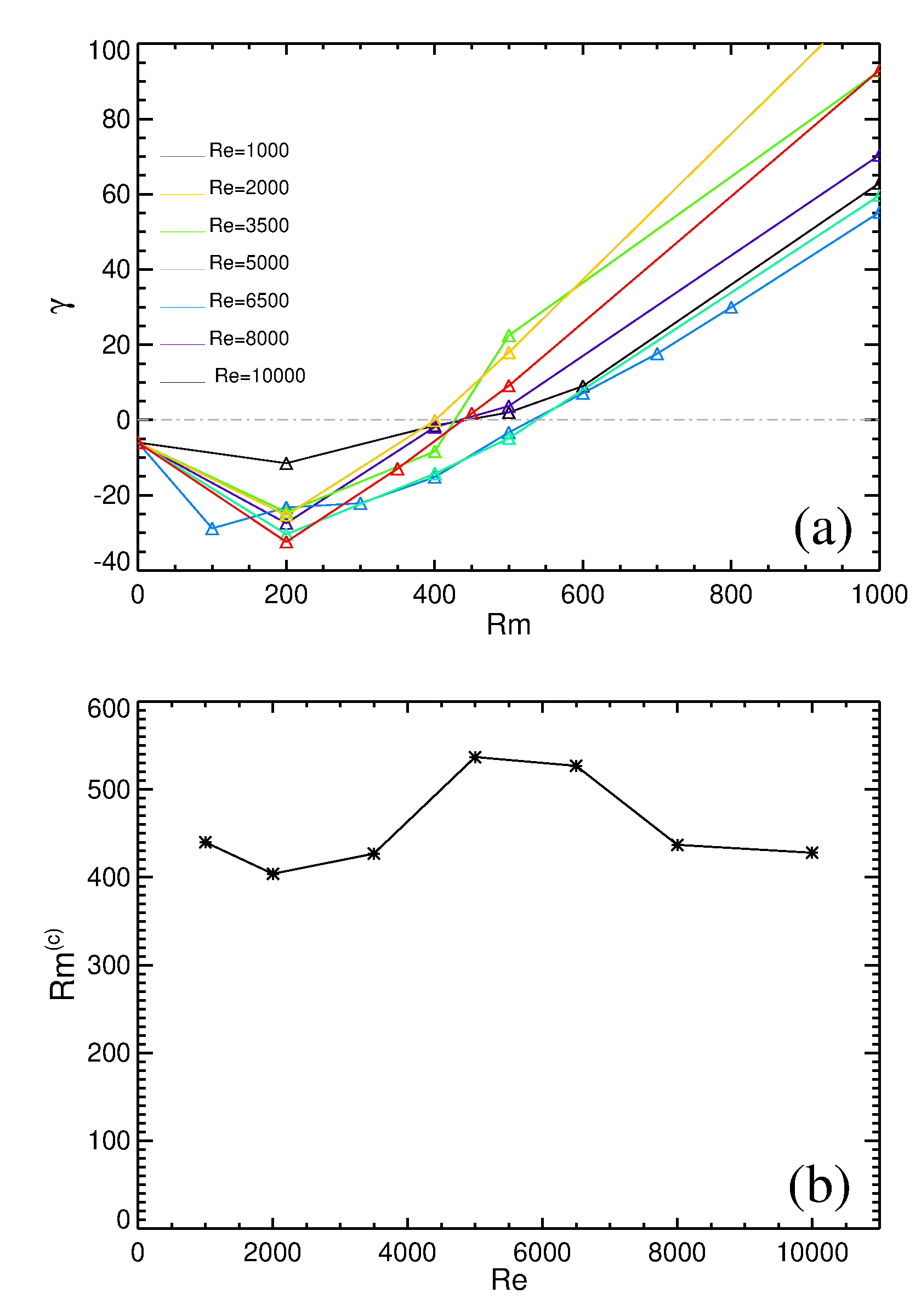}
\caption{Analysis for $\alpha=90^{\circ}$. (a) Plot of the growth rate of magnetic energy for different Reynolds number taken at the best $Po$;
(b) Critical magnetic Reynolds number in terms of dynamo action.}
\label{fig:growth_re}
\end{figure}

In this subsection we fix the nutation angle $\alpha=90^{\circ}$ to investigate the impact of the hydrodynamic Reynolds number on the flow regime and the dynamo action. We select this angle since presently it appears the best angle for dynamo action, with the lowest the critical magnetic Reynolds number.
\\ 
We start by showing the regime diagram in the $(Re,Po)$ space where the meaning of the symbols is consistent with that of Fig.~\ref{fig:phase_diagrams}: black squares denote no dynamo action, red triangles indicate dynamo action, and blue diamond signify the strongest dynamo action. The blue curve is a fit marking the scaling for the $Po^{c} \approx Re^{-1/4}$. Notice that for $Re < 3500$ we observe dynamos also significantly above the threshold curve; by contrast for larger $Re$ the dynamo action is restricted in a quite narrow range. In the next step we select the best precession ratio for every $Re$ (the blue diamonds) and show the growth rate $\gamma$ as a function of $Rm$ in Fig.~\ref{fig:growth_re}(a). The slopes of the curves seem to converge for the highest Reynolds number considered here. Collecting the points where the lines cross the $\gamma=0$ we plot the critical magnetic Reynolds number in Fig.~\ref{fig:growth_re}(b). The trend is not monotonic, showing a flat maximum in the range $4000<Re<8000$.  The smallest critical magnetic Reynolds number is found for $Re=2000$. This might be the case since at small Reynolds the flow tends to remain well organized in large scale structures rather than become turbulent with the presence of small scales.


\section{Conclusion and Prospects}

In this study, we investigated the effects of different nutation angles (60$^{\circ}$, 75$^{\circ}$ and 90$^{\circ}$) on a precession-driven flow in cylindrical geometry for both prograde and retrograde motion. We compared experimental results to direct numerical simulations. Experimentally, the axial flow $u_z$ was measured by an UDV sensor mounted on the end cap of the cylinder. These velocities were decomposed into several $(m, k)$ modes. We chose $(m, k) = (1, 1)$ and $(m, k) = (0, 2)$ modes for our study because they have significant amplitudes and are relevant for dynamo studies. 
In all cases, the experimental results agreed well with the numerical findings. For prograde precession with $\alpha$ = 60$^{\circ}$, 75$^{\circ}$ and 90$^{\circ}$ the flow abruptly transitions from a laminar to a turbulent regime, which goes along with the sudden decrease of the directly forced flow.
By contrast retrograde motion does not show a clear breakdown of the directly forced mode $(m, k) = (1, 1)$, but rather a smooth increase of the axisymmetric mode $(m, k) = (0, 2)$. 

The tendency of retrograde precession to provide a stronger large scale flow amplitude without a breakdown of the directly forced mode should be interesting for dynamo purposes, because in principle this allows an injection of more energy into the flow without breaking the base flow. 
With this question in mind, we further investigated whether the (time-averaged) flow fields obtained from the hydrodynamic simulations are capable of driving a dynamo. We conducted kinematic dynamo simulations, which can be summarized according to the following points:
\begin{itemize}
    \item The nutation angle $\alpha$ is crucial both for the hydrodynamic flow structure and the resulting dynamo action. In the phase diagram Fig.~\ref{fig:phase_diagrams} we have shown that 
    (at the present state) the most efficient 
    dynamo (with the highest growth rate $\gamma$) is found at $\alpha=90^{\circ}$. The reason for that lies in the rich and optimal flow structure for this nutation angle \cite{pizzi2021prograde,pizzi2021bl}. With view on Fig.~\ref{fig:growth_rate} it is tempting to assume that a slightly retrograde motion might provide an even lower threshold for the onset of dynamo action.
    \item For the particular case $\alpha=90^{\circ}$ the hydrodynamic Reynolds number slightly affects the best precession ratio range where dynamo is found. This precession ratio scales as $Po^{c} \sim Re^{-1/4}$. At low $Re$ the dynamos occur in a range of $Po$ more extended than for larger Reynolds, e.g $0.120< Po < 0.200$ for $Re=2000$. The critical magnetic Reynolds number shows a weak dependence on $Re$ with a slight increase around $Re \approx 6000$, but approaches the previously known value of 430 when going to  $Re=10000$. Given that the real dynamo experiment can achieve an Rm value of 700, there seems to be a reasonable safety margin to reach dynamo action.
     However the extrapolation to the hydrodynamic regime of the DRESDYN precession experiment must be considered with a grain of salt and has to be
     carefully checked in the larger experiment.
    \item The structure of the (azimuthal) magnetic field depends on the nutation angle $\alpha$, too.  
\end{itemize}

The present work can be extended in several directions.
The possibility of the down-scaled water experiment to reach Reynolds numbers of up to 2 million should be utilized to confirm the -1/4 scaling of the critical precession ratio also for nutation angles different from 90$^{\circ}$.
From the numerical point of view there is the possibility to use stress-free boundary condition for the velocity on the endcaps in order to check the specific impact of those endwall's boundary layers. 
The kinematic dynamo code should be extended to the use of vacuum boundary conditions which might still lead to some changes of the critical $Rm$ when compared to the presently used vertical field conditions.
Finally, in a more advanced study the fully coupled system of induction and Navier-Stokes equations including the back-reaction of the Lorentz forces should be investigated. For our precession system, with its very sensitive dependence on various parameters, this fully non-linear system promises to show particularly interesting effects.

\section*{DATA AVAILABILITY}
The data that support the findings of this study are available from the corresponding author upon reasonable request.

\begin{acknowledgments}
This project has received funding from the European Research Council (ERC) under the European Union’s Horizon 2020 research and innovation program (Grant Agreement No. 787544).
\end{acknowledgments}

\providecommand{\noopsort}[1]{}\providecommand{\singleletter}[1]{#1}%
%


\end{document}